\documentclass[journal,draftcls,onecolumn,12pt,twoside]{IEEEtran}
\usepackage{algpseudocode}
\usepackage{algorithm}
\usepackage{graphicx}
\usepackage{amsmath}

\usepackage{amssymb}
\usepackage{amsthm}
\usepackage{epstopdf}
\usepackage{cite}
\usepackage{rotating}
\usepackage{mathtools,lipsum}
\usepackage{mathrsfs}
\usepackage{upgreek}
\usepackage{bm}     
\usepackage{amsfonts}
\usepackage{textcomp}
\usepackage{caption}
\usepackage[numbers]{natbib}

\usepackage{booktabs,xltabular}
\usepackage{array}
\usepackage{arabtex}
\usepackage{ragged2e}
\usepackage{utf8}
\setcode{utf8}
 
\usepackage{geometry}
\usepackage{fancyhdr}
\usepackage{amsmath}
\usepackage{changepage}
\usepackage{multirow}
\usepackage{tikz}
\usepackage{xurl}
\usepackage{pstricks}
\usepackage{lmodern,tikz,lipsum}
\usepackage{enumitem}
\usepackage{colortbl}
\usepackage{hhline}
\usepackage{longtable}

\emergencystretch=\maxdimen
\usepackage{refcount}

\ifCLASSINFOpdf
 \else
\fi
\begin{document}
\addtocontents{toc}{\protect\thispagestyle{empty}}


\thispagestyle{empty}
\newpage

%
\title{Deep learning approach for  interruption attacks detection in LEO satellite networks }
%
%
%

\author{Nacereddine Sitouah, 
        Fatiha Merazka, Abdenour Hedjazi 

       \thanks{N. Sitouah,  F. Merazka and  Abdenour Hedjazi are with the Department of Telecommunications, Electrical Engineering Faculty, USTHB University, 16111, Algiers, Algeria, e-mail: \{nacereddine.sitouah@gmail.com, fmerazka@usthb.dz, abdouzpo@live.fr }
}
\maketitle
\vspace{-2cm}
\begin{abstract}
\footnotesize

The developments of satellite communication in network systems require strong and~effective security plans. Attacks such as denial of service (DoS) can be detected through the use of~machine learning techniques, especially under normal operational conditions.
This~work aims to~provide an~interruption detection strategy for~Low Earth Orbit (\textsf{LEO}) satellite networks using deep learning algorithms. Both the training, and~the testing of~the proposed models are carried out with our own communication datasets, created by~utilizing a~satellite traffic (benign and~malicious) that was generated using satellite networks simulation platforms, Omnet++ and~Inet. We test different deep learning algorithms including Multi~Layer~Perceptron (MLP), Convolutional~Neural~Network (CNN), Recurrent~Neural~Network (RNN), Gated~Recurrent~Units (GRU), and~Long~Short-term~Memory (LSTM). Followed by a full analysis and investigation of~detection rate in both binary classification, and multi-classes classification that includes different interruption categories such as Distributed DoS (DDoS), Network Jamming, and~meteorological disturbances. Simulation results for both classification types surpassed 99.33~\% in terms of~detection rate in scenarios of~full network surveillance. However, in more realistic scenarios, the best-recorded performance was 96.12~\% for the detection of binary traffic and~94.35~\% for the detection of~multi-class traffic with a~false positive rate of 3.72~\%, using a~hybrid model that combines MLP and~GRU. This Deep Learning approach efficiency calls for the necessity of using machine learning methods to improve security and to give more awareness to search for solutions that facilitate data collection in LEO satellite networks.

\end{abstract}



\IEEEpeerreviewmaketitle

\normalsize
\section{Introduction}

Nowadays, the internet has become more and more important in our lives; in~order to satisfy the increasing users' demands and their needs in term of speed, latency and~availability, we are witnessing a very competitive race of network equipment revolutions, services visualization and the emergence of the clouds. An~(artificial) satellite is an object sent into space and placed in orbit around the~Earth, it~is composed of many components including a~power system, an~altitude and orbit control system, and most importantly a communication system allowing the transmission and reception of messages \cite{1}. Satellites enable the communication between geographically dispersed systems through wireless communication links. They are widely used in various fields of communication, such as the Internet, radio broadcasting services, and telephony applications. They~allow wider coverage with reasonable deployment costs compared to~terrestrial networks. Satellite communications are expected to~become an~independent part of the fifth generation (\textsf{5G}) cellular networks and they can be the~building block for the sixth generation (\textsf{6G}) of wireless communication technologies. In \textsf{2020}, \textsf{China} successfully launched an experimental test satellite which is considered to be "The World's First 6G Satellite". In addition, satellites are making much greater use of the communication potential using Internet of~Things (\textsf{IoT}) equipment. In favor of satellites communication safety and~reliability, just like both wired and wireless terrestrial communication networks, the investigation of security must be a priority, the nature of Wide Area Network (WAN) communications, especially those that provide wider coverage through  broadcasting protocols, makes satellite networks highly vulnerable to~cyber-attacks of various types targeting large infrastructures, they typically employ very effective, and well-supported technologies to breach their targets' defense system. One of most successful and invasive attacks is the denial of service (DoS) attack. A DoS attack consists of deliberately sending a big amount of bogus messages to the operating server of the network, aiming to disrupt and satellite communication services. This attacks exhausts the limited resources of a satellite system components, which includes the memory, CPU, bandwidth and also electrical energy. Another simple, yet very efficient method of causing a DoS on~satellite networks is Jamming, just as described in its name, this attack involves the use of one or multiple devices to intentionally interfere with radio signals sent and received from and to the satellite, compromising the communication channel with the clients or other satellites.

Because the DoS attack causes very critical security issues for wireless communication networks, there must be efforts dedicated to prevent or at least remedy the damages caused by it. The available computational power of this day and age, has made artificial learning and particularly Machine Learning (ML) one of the most successful technologies for the  detection, the prevention and the prediction of security problems and incidents. We aim in this work to propose a~protection solution based on ML and demonstrate how artificial intelligence can also help ensuring the safety and availability of~non-terrestrial networks. 

The rest of this paper is structured as follows. This section will also include a state of the art regarding related works, and an overview of our contributions through this paper. In Sect. 2, an introduction to Low Earth Orbit (LEO) satellites security will be given with a brief description of some of the most important interruption detection techniques currently deployed. The detailed proposed mechanism will be presented in Sect. 3. And in Sect. 4 and 5, the environment preparation details and the performance evaluation of our solutions will be provided. And finally, a~conclusion is given in the last section of this paper.

\subsection{Related Work}

\subsubsection{Satellite communication security}

Regarding the related literature on satellite security, an efficient and secure anonymous authentication solution for satellite communication was proposed in \cite{[27]}. It is based on a simple and secure one-way hashing function to match computation capabilities in lightweight-device environments. Authors in \cite{[28]} created an access verification protocol (BAVP) using identity-based encryption and blockchain technologies. The combination makes data storage more secure in distributed computing systems, particularly LEO satellite networks. The simulation proved the security of the protocol but the deployment still needs some reforming of the blockchain technology according to~particular satellite network routing algorithms. Another authentication schemes for~satellite-communication using lightweight key agreement was proposed in \cite{[29]} \cite{[30]} and \cite{[31]}. Authors of \cite{[29]} provided mutual authentication, session-key agreement and~a~solution for user anonymity problems. They presented formal and informal security analyses against different well-known attacks. Their results validate the~new scheme efficiency and additional security features such as user anonymity and~forward secrecy. In  \cite{[32]}, they treat jamming threats to enhance security, the~authors performed a secrecy analysis for the uplink transmission in satellite communication which is based on the orthogonal time-frequency space (OTFS) and a comparison with the traditional Orthogonal Frequency Division Multiplexing (OFDM) scheme. Results demonstrated better performance of the uplink LEO Sat-Com (Satellite Communication) system.
\subsubsection{Machine learning in satellite communication} 
The state of the art concerning satellite communication security indicates that non-intelligent methods, which do not include any type of artificial intelligence, significantly underperform compared to solutions integrating intelligence. 
The work in \cite{[33]} puts forward an Extreme Learning Machine-based distributed routing (ELMDR) strategy to make routing decisions by predicting traffic. In \cite{[34]}, they use machine learning algorithms to~detect message collisions, by optimizing the automatic dependent surveillance-broadcast (ADS-B) reception. The NASA research center and authors of \cite{[35]} studied the role of machine learning in the link-to-link aspect of Sat-Com, they used reinforcement learning space links to ground stations using NASA's testbed on the international space station. In \cite{[36]} authors created a lightweight model of neural networks in the hope of encouraging the development of an intelligent edge IoT architecture by satellite, this solution combines edge computing with deep learning. Results show and prove that lightweight neural networks like MobileNet and ShuffleNet are more suited to satellite IoT scenarios in the IoT edge satellite intelligent computing architecture. The work in \cite{[37]} experiments on the promising transport protocol  Multipath TCP (MPTCP) in~LEO Sat networks, using the self-learning properties brought by reinforcement learning in order to find the optimal congestion control strategies.
 Another potential of improvement by the use of deep learning techniques is the accuracy of~5G terminal location discovery. \cite{[38]} overcomes the dependency on the external global navigation satellite system, data clustering methods of unsupervised ML are~used to classify and exclude measured data. \cite{[39]} emulates a Sat-Com platform to generate environment data and capture different internet communications (encrypted, unencrypted and tunneled) in order to~have a high classification rate and improve the overall Quality of Service (QoS) of~the communication. Machine learning is also recently used in satellite operation, such as interference detection, flexible-payload configuration and congestion prediction \cite{[40]}, and for resource allocation on the internet of Remote things \cite{[41]}, the authors study different security aspect in SIoRTNs (satellites for Internet of Remote things networks), the~proposed algorithm based on actor-critic reinforcement learning (SACRL) showed effectiveness in term of the IoRT data download performance. 
\subsubsection{Detection of interruption using machine learning }
DDoS and interruption attacks are the biggest threat that generally target Sat-Com. We performed a~comprehensive literature review regarding these particular issues. The first obstacle was the unavailability of~communications datasets, followed by the unavailability of~public DDoS traffic captured from satellite networks. However, our work applies the same principle of using ML for DDoS detection as in-ground networks. Experiments in \cite{[42]} use public communication datasets to train a model capable of detecting several types of attacks including the DoS in the IoT, by~deploying a distributed learning method based on Fog Computing.\\ Authors of \cite{[43]} proposed a solution based on artificial neural networks for the recognition of DDoS attacks, combined with a Learning Vector Quantization neural network (LVA NN). The authors of \cite{[44]} are interested in the security of the IoT, particularly the IIotT (Industrial IoT), they developed an anomaly detection technique for the IICSs (Internet Industrial Control Systems)  based on ANNs models that can learn and validate using information collected from TCP/IP packets. Similarly, researchers in \cite{[45]} studied the detection of DDoS attacks in~smart agriculture,  called AIoT.  They deploy an identifier based on three models (CNN, DNN, and RNN) and compare its performance for both binary  and~multi-class classifications. Another work in  \cite{[46]}  developed a model based on~CNNs, with both types of classification, using the most recent IDS datasets that~contains advanced DoS attacks. \cite{[47]} took a different approach using a hybrid deep learning method for Botnet attack detection in the IoT, the method overcomes memory constraints in LSTM models, authors used a short-memory automatic encoder (LAE) to reduce feature size, and then implemented Bi-LSTM for~interruption attacks detection. 
\subsection{Contributions}
In this paper, we propose a methodology that allows the study and the investigation of deep learning solutions in the security of satellite communication. Our main contributions can be expressed into the following: 
\begin{itemize}
    \item [\textbullet] We present a simulation environment which allows the study on similar characteristics of real satellite networks, using \textsf{Omnet++ network simulator}, and a personalized topology for four different scenarios: Normal communication, DDoS flooding attacks, Natural cause interruptions (ex: rain and thunderstorms) and network jamming.
    \item [\textbullet] We propose a method to generate rich datasets from the simulation results, then we present different deep learning algorithms and experiment with them on our datasets in different scenarios and conditions.
    \item [\textbullet] We test each of our algorithms on the two main datasets:  SAT-COM.LEO.NDBPO\#1 and SAT-COM.LEO.NDBPO\#2, for both classification types:  binary classification and multi-class classification in an offline manner.
    \item [\textbullet]  We finally present a methodology that allows the test of the previous algorithms combined with probability calculations to create an IDS capable of real-time detection of interruption causes.
    \item [\textbullet] The performance of each algorithm in each scenario is investigated and~compared to ensure efficiency and minimize false alerts for better overall performance. 
    
\end{itemize}
. 
  \newpage
\section{LEO security }
 
 China's CCID (China Center for Information Industry Development) estimates that the number of LEO satellites will exceed 57,000 by 2029 \cite{[20]}
 , encompassing all technology sectors. Hence, any damage inflicted in the satellite sector can have a quality of service effect, leading to heavy financial or data losses. Unfortunately, organizations rarely have direct control over satellites cybersecurity, as satellite operations are driven by technologies hosted on Earth; these land-based entry points offers cyber attackers a huge number of potential forays.
 Long range telemetry for communication with ground stations poses a great weakness, uplinks and downlinks are often transmitted via open telecommunication network security protocols, and easily accessible by cybercriminals. IoT devices that use satellite communications are also a potential entry points for bad actors.
\subsection {Classification of security problems}

According to the work of \textit {\textsf {H. Cao et al}} \cite{[21]}
, security problems in satellite communications can be divided into three broad categories; we focus on  telecommunication and network security:

\begin {itemize}
    \item [\textbullet] \textsf {National security}
    
\textit { {1 - Threats to national security
 :}}
 These threats are summed up in the theft of~strategic information on targeted countries, by deploying an Earth observation station on LEO satellites, it's an eavesdropping attack condemned by the \textit{\textsf{Space Law}}.
   
\textit { {2 - preemption of frequency and orbit resources
 :}}
The ITU organizes and~shares orbit and frequency resources according to a FIFO service queue, this type of~threat consists of the illegal use of these limited resources, which can potentially cause an interruption or a \textsf{DoS}.

\item [\textbullet] \textsf {Network security}

\textit { {3 -
Identity theft:}} If the authentication and identification mechanism used is~vulnerable, old or public, an attacker can impersonate a legitimate terminal and gain access to it by calculating uplink frequencies and then bypassing ACPs (Access control policies), or a spy satellite can invoke an ISL connection that seems legitimate with the victim satellite.

\textit { {4 - Listening to data and intercepting information:}} Many satellite communications do not encrypt the data transfer, and most protocols are~either unscalable or hard to upgrade, which makes  listening and~intercepting messages very frequent and easy to execute.
 
\textit { {5 - signal interference:}} This kind of attacks is the most common and efficient, an attacker can interfere with satellites and scramble channels by transmitting signals from transmitters with  higher power, on the communicating frequency band.

\textit { {6 - denial of services:}} Like terrestrial networks, \textsf{DDoS} are effective in attacking Internet Satellites, an attacker simulates a flow of legitimate and bogus requests from satellite terminals in order to send it to satellites, preventing legitimate terminals from good service quality.

\textit { {7 - Malicious occupation of satellite bandwidth resources:}} Most satellites operate as a transponder without unpacking the signal, so it becomes impossible to verify the legitimacy of a received signal, even if the satellite manages to unwrap the signal, the attacker uses his own cryptography technique to hide his communications.

\item [\textbullet] \textsf {Equipment safety}
    
\textit { {8 - Malicious satellite control:}} An attacker sends malicious instructions or~injects viruses into satellites from the ground or from space to achieve the objective of controlling the satellites, dragging the satellite out of orbit, interrupting services or contaminating other satellites.

\textit { {9 - Malicious consumption of satellite resources:}} An attacker directly affect the~lifespan of satellites by consuming thrusters (responsible for satellite movement, maintaining orbit and avoiding collisions).
\end {itemize}


\subsection{Interruptions detection techniques}

\textsf{An interruption} is a temporary shutdown or a temporary unavailability of~a~satellite network supply to a customer, such as signal interference, denials~of~service, unauthorized use of resources, malicious network jamming. The~protection against such attacks can be classified into four methods according to  \cite{[20]}:
\begin {itemize} [leftmargin = -0.3em]
\item [\textbullet] \textsf {Statistical-based methods:} Based on various statistical algorithms developed using different measures. Detection begins with data collection, and ends with an analysis of a significant sample of the network to identify malicious traffic. \\In \cite{[22]}
, a detection based on the entropy anomaly is proposed, it can determine whether the current state is normal or compromised based on the entropy value, then detects the DDoS attack by changing the entropy value of the network characteristics. Moreover, Shannon's entropy \cite{[31]} 
is considered to be one of the best methods to detect abnormal traffic \cite{[23]} \cite{[24]}.
 \item [\textbullet] \textsf {Machine learning method:} Traditional ML approaches are used to analyze and classify traffic. They obtain good results under certain conditions. In \cite{[25]}
 , a method based on K-means and K-nearest fast neighbors is proposed. The limitation of traditional machine learning for detecting DDoS attacks is that historical traffic characteristics cannot be used, and the bigger the data grows, the lower the detection rate. It is for this reason that new ML approaches such as DL give more hope for better results.
 \vspace {0.2cm}
 \item [\textbullet] \textsf {Prevention techniques:}
One of these works is presented in \cite{[26]} 
in which the authors studied the possibility of preventing DoS attacks by reducing the power consumption of the control center while preventing DoS attacks. others propose a~system that provides proactive prevention against DoS and DDoS attacks, monitoring the network in normal operation with the average number of requests circulating, making it  possible to block any traffic generating an anomaly. In~general, implementing any authentication and cryptography solution dramatically decreases the risk of being attacked, but the majority of satellite communication systems lack investment in~security and protection techniques.
\vspace {0.2cm}
\item [\textbullet] \textsf {Specific applications:} These solutions developed and intended in particular for security against DDoS attacks, from a specific structure, such as the use of the blockchain to~advise other users on a attack, honeypots using the network to deceive the intruder.

\section{Proposed interruption detection mechanism}

 Our work  puts into practice different techniques of deep learning in order to propose a solution that improves the detection of interruption attacks targeting LEO satellite communications.
The effectiveness of any machine learning model depends on the availability of rich training data. Thus, communication datasets are the essential element of our DL based project. To exceed the problem of public DDoS traffic datasets unavailability had to be solved, we propose a simulation technique that will allow us to generate a benign communication dataset, with~several communication scenarios affected by interruptions in satellite networks, a~technique dedicated particularly to networks via LEO satellites. The well-known simulation tools  \textsf{OMNeT++} and \textsf{INET} library will be used to capture the different properties of the simulated satellite communications.
We propose a satellite networking topology composed of 20 earth terminals, three satellites equipped with processing units, inter-satellite links, and satellite-terrestrial links. Each terminal in the topology is assumed  able to communicate directly with satellites in its reception scope.

Fig.\ref{Figure: Topology of the simulated network} represents the topology used to generate satellite communication traffic during simulation, where each orbiting satellite covers an area of the earth. \textsf{Terminals{$_{1..10}$}} are covered by \textsf{Satellite{$_1$}} and \textsf{Terminals{$_{11..20}$}} are covered by the \textsf{Satellite{$_2$}}. The communication between an earth terminal and a satellite is on the frequency channel \textit{1616Mhz}. When an \textsf{End-User} from \textsf{Zone{$_1$}} wants to communicate with another \textsf{End-User} from \textsf{Zone{$_2$}}, \textsf{Satellite{$_1$}} uses the frequency channel  \textit{23180MHz} to send the message to the \textsf{Satellite{$_3$}}, which in turn transmits  it to \textsf{Satellite{$_2$}}, then this latter finally transmits the message to the destination \textsf{End-User}.
Dynamic \textit{Satellite-to-Satellite routing protocols} are not supported by the simulator, therefore, \textsf{Satellite{$_3$}} uses static routing protocols assuming that both \textsf{Satellite{$_1$}} and \textsf{Satellite{$_2$}} are fixed objects from its point of view.

\begin {figure} [! h]
\begin {center}
\includegraphics [scale = 0.9] {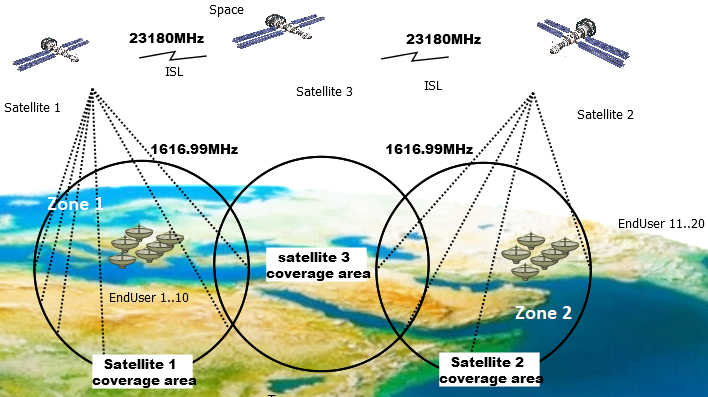}
\caption {Network topology}
\label {Figure: Topology of the simulated network}
\end {center}
\end {figure}


\subsection {Construction of <<SATCOM.LEO.NDBPO.\#1>> dataset}  We would like to mention the work of \textit{F.Alhaidari} and \textit{A.Alrehan}, although this project took a much different path, the idea of generating this dataset was inspired by their efforts in \cite{[48]}
. Essentially, two utilities play an important role for datasets' generation. First, the \textsf{log console} which is window available under the GUI of~\textsf{Omnet++}, displays message transmission events that have taken place between different modules during simulation. Second,
the \textsf{vector file}, which records network information in time intervals as statistics for each module, as data timestamp. These data values are recorded and captured based on several categories or~features.

By combining the data retrieved using both utilities, we obtain relevant information characterized with  LEO satellites communications properties. We use Jupyter Notebook with our python scripts that explore each event in the log file, and for each event it calculates the \textsf{current}, \textsf{previous} and \textsf{next time} for each module. To extract the values of the features in the interval between the \textsf{previous time} and the \textsf{next time}, the script calculates the value of each feature in the Vector file for each communication event according to the following equation \ref{equation: calculate features}:

 \textrm {\quad\quad\quad For each feature \textit {i}  and event \textit {j} (with \textit{n} the number of nodes in communication):}
\begin {equation}
ValFeat_i (event_{{j}}) = \sum_ {k = 1} ^ {n} ValFeat_i Node_k [\textrm {Previous event time, Next event time}] \label{equation: calculate features}
\end {equation}

Fig.\ref{Figure :Flux de création des fonctionnalités du dataset} presents the workflow of the dataset's creation.

\begin{figure}[htb]
\begin{center}
\includegraphics[scale=0.6]{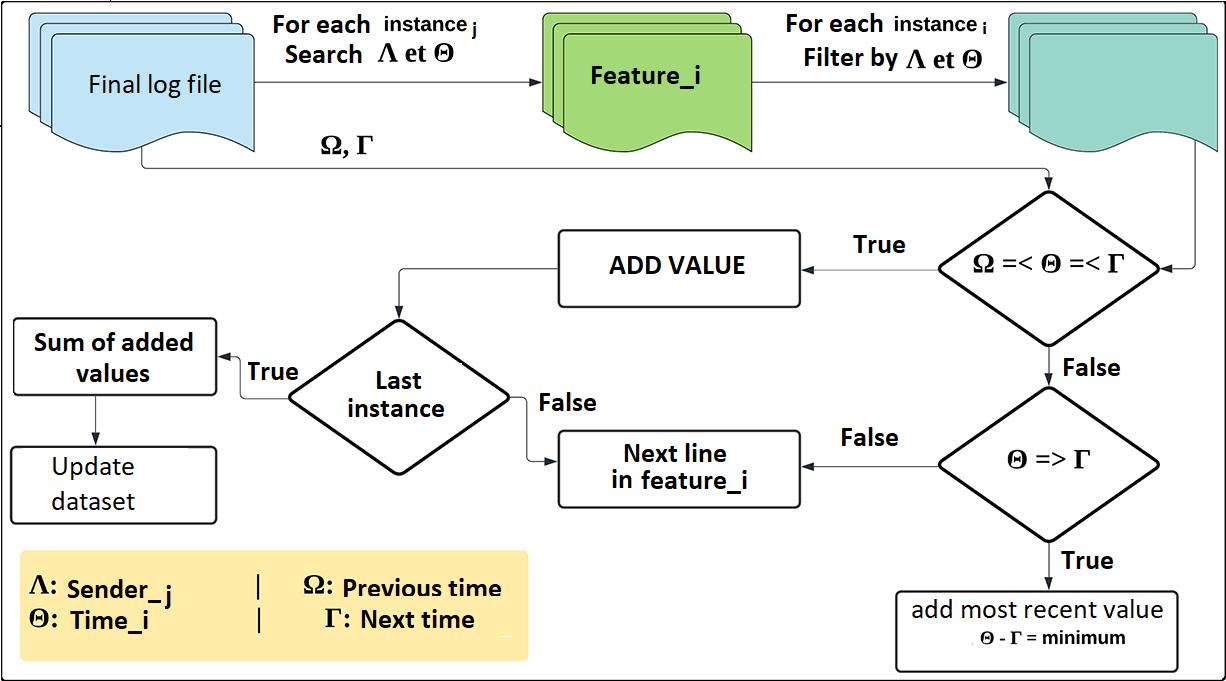}
\caption{Features extraction flow.  }
\label{Figure :Flux de création des fonctionnalités du dataset}
\end{center}
\end{figure}

We use another python script to generate more features and enrich the dataset, as~notation we call reference \textsf{local time} for packets with the same sender, receiver and~protocol, and \textsf{global time} to reference any packet sent or received.  

\subsection{Construction of <<SATCOM.LEO.NDBPO.\#2>> dataset}
This dataset serves the purpose of  implementing a flow based intrusion detection system (IDS), which would be more suitable for real-time detection. We use the module <<PcapRecorder>> of Omnet++ to record the communication traffic that passes to or from a node in the topology. After the generation of the communication pcap files we use <<TcpReplay>> tool to regenerate the simulation traffic on an   environment external to the simulation.

Several features similar to the previous dataset are used, combined with  other new  flow-based ones that are created with python scripts. A communication flow is any continuous and uninterrupted communication, that does not contains any error message due to an unreachable destination. 
\newpage
Tab.\ref{Description des caracteristiques du dataset construit} summarizes different features in~the datasets:

\scriptsize 
\begin{longtable}{|p{0.5cm}|p{3.1cm}|p{6.5cm}|p{1.5cm}|p{1.5cm}|} 

\caption{Description of \textsf{<<SATCOM.LEO.NDBPO.\#1>>} and \textsf{<<SATCOM.LEO.NDBPO.\#2>>} features}

\label{Description des caracteristiques du dataset construit}
\endfirsthead
\endhead

 \cellcolor[HTML]{B2BEB5}No. &\cellcolor[HTML]{B2BEB5}Feature &\cellcolor[HTML]{B2BEB5} Description &\cellcolor[HTML]{B2BEB5} Dataset\#1 &\cellcolor[HTML]{B2BEB5} \Centering{Dataset\#2} \\
\hline
- &	sendTime & Sending time according to the sniffed packet &  \Centering yes & \Centering  yes\\
- &	sender &  Packet sender & \Centering yes &  \Centering yes\\
- &	reciever & Packet receiver & \Centering yes & \Centering yes\\
- &	IP\_src &  Ip Address of the source &  \Centering yes & \Centering yes\\
- &	 port\_src & Port of the source &  \Centering yes & \Centering yes\\
- &	IP\_dest &  Ip Addres of the destination & \Centering yes & \Centering yes\\
- &	port\_dest & Port of the  destination & \Centering yes & \Centering yes\\
- &	Frequency &  Transmission frequency & \Centering yes & \Centering no\\\hline
1 &	Next\_Current\_diff & Global time difference between current and next  & \Centering yes & \Centering yes\\
2 &	 Next\_Pre\_diff & Global time  difference  between  previous and  next & \Centering yes & \Centering yes\\
3 &	SNext\_Current\_diff & Local time difference between current and next  & \Centering yes & \Centering yes\\
4 &	SNext\_Pre\_diff & Local time difference between previous and next & \Centering yes & \Centering yes\\\hline
5 &	size & Packet size & \Centering yes & \Centering yes\\
6 &	channel & Frequency channel & \Centering yes & \Centering yes\\
7 &	duration & Transmission duration & \Centering yes & \Centering no\\
8 &	 packet\_type & Packet type (Udp,Icmp) & \Centering yes & \Centering yes\\\hline
9 &	rcvdPK & Recieved packets& \Centering yes & \Centering no\\
10 & sentPK	 & Sent packets & \Centering yes & \Centering no\\\
11 & droppedPKWrongPort & Dropped packets, wrong ports & \Centering yes & \Centering no\\\
12 & DataQueueLen	 & Data queue length & \Centering yes & \Centering no\\\
13 & passedUpPk	 & Passed up packets & \Centering yes & \Centering no\\\
14 & rcvdPKFromHL	 & Received from higher layer & \Centering yes & \Centering no\\\
15 & rcvdPKFromLL & Received from lower layer & \Centering yes & \Centering no\\\
16 & sentDownPK	 & Sent down packets & \Centering yes & \Centering no\\\
17 & DropPKByQueue	 & dropped packets from queue & \Centering yes & \Centering no\ \\
18 & snir & Signal-to-interference-plus-noise ratio & \Centering yes & \Centering yes\ \\
19 & throughput &  Transmission throughput & \Centering yes & \Centering yes\\\hline
20 & Flow Bytes\_s	 & Flow's bytes per second & \Centering no & \Centering yes \\
21 & Flow Packets\_s & Flow's packets per second & \Centering no & \Centering yes\\
22 & meanT\_b\_2P	 & Mean time between two packets & \Centering no & \Centering yes\\
23 & maxT\_b\_2P & Maximum time between two packets & \Centering no & \Centering yes\\
24 & minT\_b\_2P	 & Minimum time between two packets & \Centering no & \Centering yes\\\hline
\end{longtable}






\normalsize

\subsection{Description of the different scenarios}
We focus on various security aspects related  to services' availability, in order to achieve this we do not consider network flooding attacks solely, we rather include meteorological and other jamming disruption scenarios.

\subsubsection {Scenario 1 <<Benign traffic>>}

Based on the topology shown in Fig.\ref{Figure: Topology of the simulated network}, benign communication is achieved by a stream based on UDP protocol, it simulates real-time running applications such as voice and video, which cannot wait for recovery mechanisms such as retransmissions. The protocol's performance through a satellite network is characterized as having a significant delay depending on the height of the satellite orbit and the number of satellite hops. It should be noted that \textsf{Omnet++} does not allow three-dimensional simulation, for this reason the height of the satellites is ignored, but the characteristics of the network are based on the actual performances of a LEO satellite. Benign traffic is labeled \textsf{<<Normal>>}.

\textit {Note}: \textsf{<<Omnet++>>} does not implement frequency division protocols, but allows communication through channel numbers, so if two nodes use two different channels, they cannot communicate.
\par \textit {Note 2}: The \textit{<<Simulation time>>} intervals can be set freely depending on the power of the computer in hand.

\subsubsection {Scenario 2 \textit{<<Malicious "UDP Flood" traffic>>}}

This traffic is a DDoS attack targeting the satellites communications, this attack is carried out by infecting certain earth terminals of the network, to send larger packets with a higher transmission rate. This malicious traffic is labeled \textsf{<<UDP\_Flood\_attack>>}, and~carried out  by adding another communication flow sent by contaminated users. Fig.\ref{Figure : Topologie du réseau attaqué <<UDP Flood>>}  is a graphic demonstration of this attack.

\begin{figure}[htb]
\begin{center}
\includegraphics[scale=0.9]{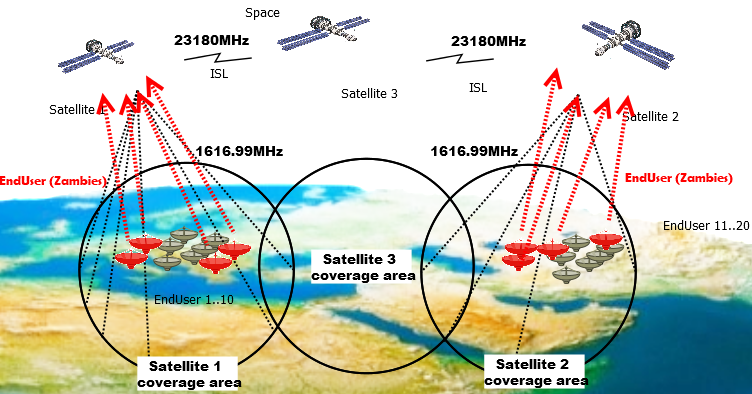}
\caption{ Network topology of UDP flood attacks.}
\label{Figure : Topologie du réseau attaqué <<UDP Flood>>}
\end{center}
\end{figure}



\subsubsection {Scenario 3 \textit{<<Natural interruption "Rain and thunderstorms">>}}
Among the natural disruptions are meteorological events, such as torrential downpours and~aggressive thunderstorms, which actually dramatically increase the rate of loss in satellite communication, This traffic is carried out by modifying  the \textsf{path loss} for the affected users. Fig.\ref{Figure : Schema du scénario 3} describes this scenario.

\begin{figure}[!h]
\begin{center}
\includegraphics[scale=0.65]{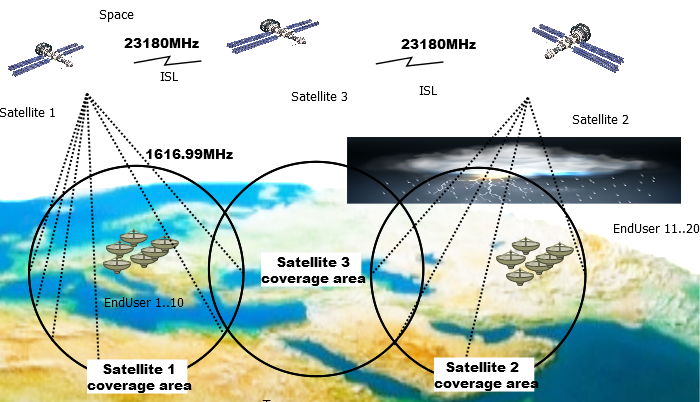}
\caption{ Network topology of natural disruptions. }
\label{Figure : Schema du scénario 3}
\end{center}
\end{figure}


\subsubsection {Scenario 4 \textit{<<Jamming network interference>>}}
We achieve satellite network jamming by simulating aircrafts \textsf{(JamCrafts)} that fly in the field of view of satellites antennas, send and receive noise signals in the same radio frequencies used by the targeted satellite, this traffic is carried out by adding another communication flow between attacking aircraft and ground agents. Fig.\ref{Figure: Topology of the attacked network <<jam>>}  is an abstraction of this attack.

\begin{figure}[!h]
\begin{center}
\includegraphics[scale=1]{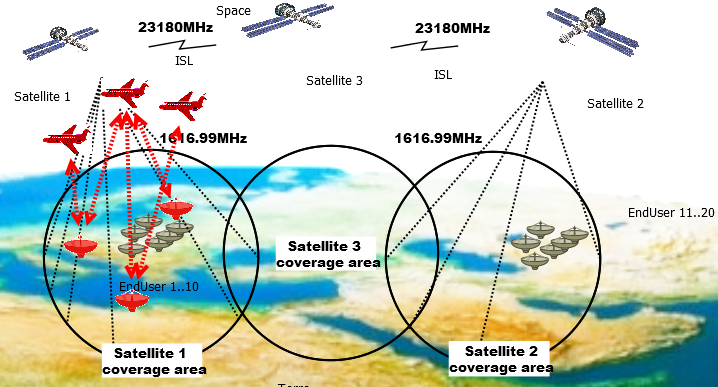}
\caption{ Network topology of jamming attacks.}
\label{Figure: Topology of the attacked network <<jam>>}
\end{center}
\end{figure}

\newpage
Tab. \ref{Description des caracteristiques du trafic bénin} Describes the characteristics of different flows and scenarios.

\begin{table}[!h]
\scriptsize
\caption{Description of the characteristics of benign traffic }
\label{Description des caracteristiques du trafic bénin}
\centering
\begin{tabular}{|p{5.5cm}|p{2cm}|p{2cm}|p{2cm}|p{2cm}|} 
\hline
 \cellcolor[HTML]{B2BEB5}Parameters &  \cellcolor[HTML]{B2BEB5}Scenario 1 & \cellcolor[HTML]{B2BEB5}Scenario 2 & \cellcolor[HTML]{B2BEB5}Scenario3 &  \cellcolor[HTML]{B2BEB5}Scenario 4\\
\hline \hline
Number of terminals &  20 &  20 &  20 &  20 \\
 Number of affected terminal & / &  6 & 10 & /\\
Number of satellites &  3 &  3 &  3 &  3\\
Benign source ports & 5555,3099,2099 & 5555,3099,2099 & 5555,3099,2099 & 5555,3099,2099 \\
Benign destination ports & 2000,9901,9902 & 2000,9901,9902 & 2000,9901,9902 & 2000,9901,9902 \\
Targeted satellite & / & satellite 3 &  satellite 2 & satellite 1  \\
Malicious source ports & / & 2001 & / & / \\
Malicious destination ports & / & 2002 & / & / \\
Total number of channels & 22 & 22 & 22 & 22\\
Normal Packets size & [40B,635B] & [40B,635B]  & [40B,635B] & [40B,635B]\\
Attack Packets size & / & [4000B,5000B]  & / & /\\
Normal transmission rate & [100ms,400ms]& [100ms,400ms]& [100ms,400ms]& [100ms,400ms] \\
Attack transmission rate &  / &[20ms,50ms] & / & /\\
Inter-satellite communication channel & 30 , 31 & 30 , 31 & 30 , 31 & 30 , 31  \\
Ch between satellite 0 \& EndUsers [0..9] &  [0..10] &  [0..10] &  [0..10] &  [0..10] \\
Ch between satellite 0 \& EndUsers [10..19] &  [10..19] &  [10..19] &  [10..19] &  [10..19] \\
Path loss & 2 & 2 & \textbf{4} & 2 \\
Satellite transmitter power & 7W  & 7W  & 7W  & 7W\\
EndUser transmitter power & 7W  & 7W & 7W & 7W \\
Number of ground agents &  / &  / &  / & 10 \\
Number of  JamCrafts &  / &  / &  /  & 1 \\
Ch between JamCraft \& JamUsers [0..9] &  / &  / &  /  &  [0..10] \\
JamCraft   transmitter power &  / &  / &  / & 12W \\
JamUser transmitter power &  / &  / &  / & 20W \\
Attack duration & 0s  & 33s & 26s & 80s \\
Simulation time (dataset\#1) &  0s-90s & 90s-123s & 124-250s & 250s-330\\
 Simulation time  (dataset\#2) & 0s-900s  & 900s-1500s  & 3000s-4500s  & 1500s-3000s \\\hline
\end{tabular}
\end{table}



\subsection {Applying Deep Learning}

The first step in applying a machine learning algorithm is data normalization, which is the process of resizing the dataset's attributes into a particular range, such as between \textit{0 \& 1} or \textit{1 \& -1}. Data normalization prepares datasets to be fed into ML classifiers, in order improve the accuracy of the results. We normalize our datasets with the \textsf{min-max} function \cite{[49]}:
\begin {equation}
    X = \frac {(x-Min)} {(Max-Min)}
\end {equation}
\\
We then standardize and label the datasets, and implement different deep learning algorithms and compare their efficiency, we  test several models such as MLPs, CNNs and RNNs, the workflow to achieve our goals is as follows:
\subsubsection {Training phase}
This is the first step to create a classifier, it is necessary to find the~most adequate parameters of the algorithms for the classification problem in~hand, this makes it possible to create an intelligent model offering high precision rates. The~input dataset dataset will be divided into three parts containing 40\%, 30\% and~30\% of the dataset's totality, the first part is used to train the model and modify the weights of the algorithm. To ensure that the model does not overfit, the~second part is used for the validation of the model, so that at each iteration, the~model is~only updated if it performs better on the validation set.

\subsubsection {Evaluation phase}
The last 30\% of the splits, is used to test the new model, and~display  results in a format allowing its evaluation, which is done according to~the following factors
(\textit{Annotations: }{TP} = \textit{True positives}, {FP} = \textit{False positives}, {FN}~=~\textit{False~negatives}, {TN} = \textit{True negatives} \cite{[49]}):
 \begin {itemize}
\item [\textbullet] \textsf {Accuracy}: The percentage of correct classifications of the model:\\
$  {\textit {Accuracy}} = \frac {TP + TN} {TP + FP + TN + FN} $

\item [\textbullet] \textsf {Precision}:  The percentage of correct positive instances out of the total predicted positive cases:
$  {\textit {Precision}} = \frac {TP} {TP + FP} $
 
\vspace {0.2cm}
\item [\textbullet] \textsf {Recall}:  The percentage of positive instances out of the total of actual positive instances:
$  {\textit {Recall}} = \frac {TP} {TP + FN} $

\vspace {0.2cm}
\item [\textbullet] \textsf {Sum of probabilities}:  The total sum of the probabilities of the correct classes:
$  {\textit {sum\_prob}} = \sum_ {k = 1} ^ {n} Prob \_class \_correct $

\item [\textbullet] \textsf {Confusion matrix}:  The rate of false or correct predictions by class.
\end {itemize}

\subsection {Workflow/Different steps of implementation}
\subsubsection {Offline detection under simulation}
Here, we suppose that the controller (agent), has full surveillance on the simulated LEO network, so we assume he can monitor all~the~modules participating in the communication, this means that the generated dataset contains all the packets sent or received, and  all features related to the network. 
This way, it becomes possible to find the most significant characteristics and have a precise and deeply detailed description of the important element for~a~safe satellite network (in the simulation environment).
 
\subsubsection {Offline / online detection} :
In this scenario, the specific constraints and conditions of a satellite communication network are taken more into account, such as: Electric power limitation, Computing power constraints, information availability and unavailability, Necessity of online detections, Impossibility of live surveillance on \textsf{EndUsers}.

 In this case, we set surveillance at satellites level, the security agent can only monitor them or earth stations. This makes it possible to simulate a satellite by~a~virtual machine which will capture in real time, the  communication flow generated from the simulation environment. This approach was used to create the~\textsf{<<SATCOM.LEO.NDBPO.\#2>>} dataset. \par
\textsf{TcpReplay} tool allows us to generate traffic with the same send and receive time values, the satellite uses our own sniffer programmed in python, that captures the~packets, processes them and transmits the processed data via the best performing deep learning model.

\section{Environment preparation and datasets generation}
In this section, we put into action our solution, evaluate the generated flow and~our~anomaly based detection approach in a working environment that allows the~creation of a communication network composed of LEO satellites.
\subsection {Work environment}
We use an ultra-portable Notebook "Lenovo T490s" with a Core (TM) i7-8665U processor (1.90 GHz, up to 4.80 GHz with Turbo Boost, 4 Cores, 8 Threads, 8 MB Cache), 64 bits and 16 GB of RAM, equipped with a Windows 10 Pro operating system and two VMs with a Linux OS Ubuntu 16.04 LTS.

Installation of the simulator:
We install the \textsf{Omnet++} simulation tool, its vendors offer a step-by-step manual which can be found under \cite{[50]}. 
 We use \textsf{Omnet++ v4.6} but we recommend installing it under a \textit{Linux environment} for version compatibility reasons, we use \textit{Ubuntu 16.04}. After installing the simulator, it~is~necessary to download and import the open source \textsf{INET v2.5} model library into \textsf{Omnet++}, a complete manual can be found under \cite{[51]}.

Topology implementation:
To achieve the architecture seen in Fig.\ref{Figure: Topology of the simulated network}, we use a~combination of modules created under \textsf{Omnet++}, detailed in Tab.\ref{desctiption of the main omnet  nodes.}:
\begin{table}[!h]

\scriptsize
\caption{Description of the main \textsf{Omnet}'s nodes in the network topology }
\label{desctiption of the main omnet  nodes.}
\centering
\begin{tabular}{ |p{2.2cm}|p{12cm}| } 

\hline
 \cellcolor[HTML]{B2BEB5}Name &  \cellcolor[HTML]{B2BEB5}Description\\
\hline \hline
 EndUser & An extension of Inet's 'Sat\_User' module modified to represent a terrestrial user.\\ \& JamUser &  \\\hline 
\textcolor{white}{....................} Satellite & An extension of the 'Satellite' module of OS3 (Omnet's Open Source Satellite Simulator developed at the Communication Networks Institute, TU Dortmund, Germany \cite{[58]}) modified and equipped with a~communication network interface and~a~radio interface to represent a~satellite able to~receive and transmit messages. \\\hline
\textcolor{white}{............. }ChannelControl & This is an instance in every network model that contains mobile or wireless nodes. This module is informed of the location and movement of nodes, and determines which nodes are within communication or interference distance. \\\hline
   JamCraft & A custom module representing an aircraft with the ability to transmit stray radio signals covering a large land area. \\\hline 

\end{tabular}
\end{table}


\begin{figure}[!h]
\begin{center}
\includegraphics[scale=0.65]{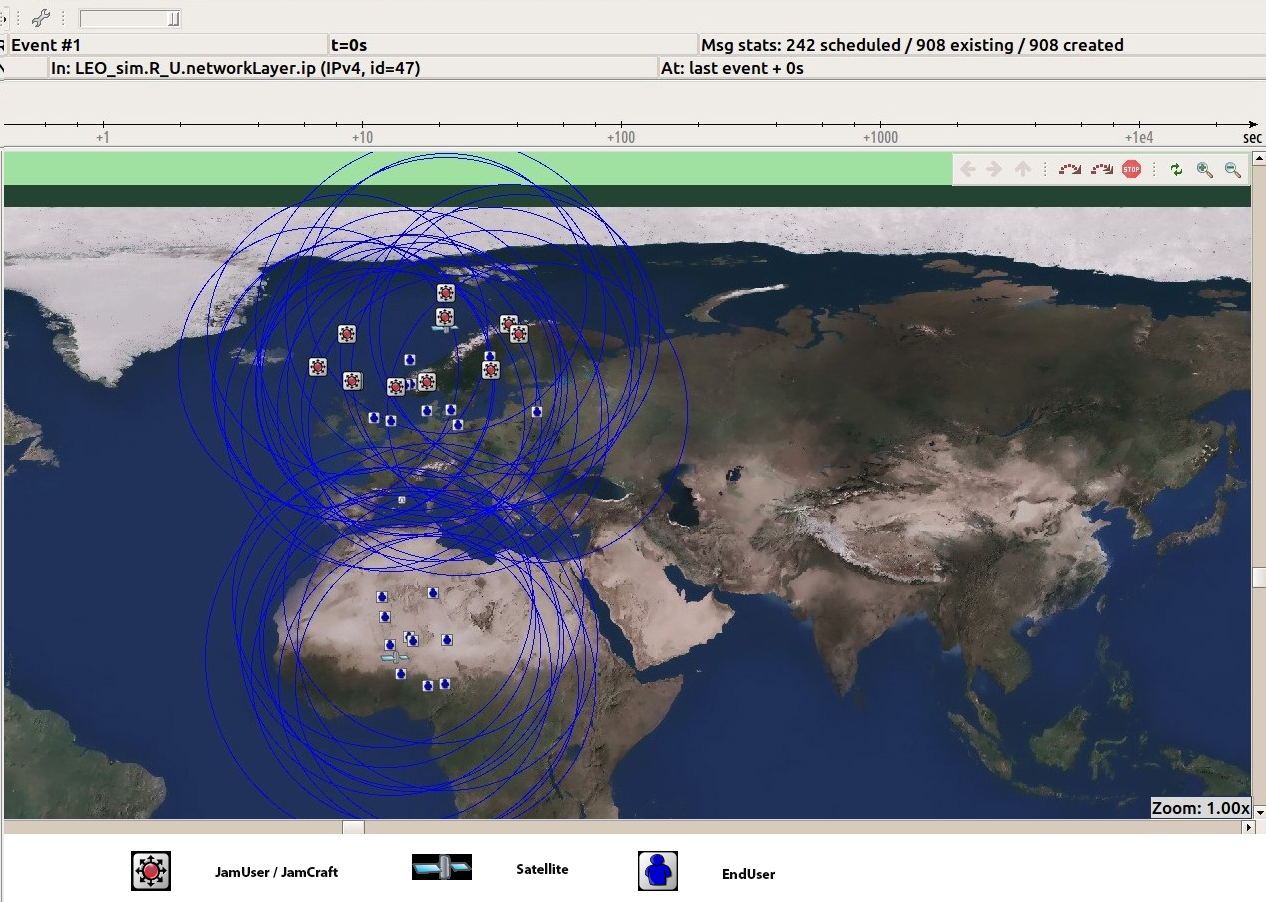}
\caption{ Topology under Omnet. }
\label{Topologie sous Omnet}
\end{center}
\end{figure}

\subsubsection{Dataset generation}
We generate \textsf{<<SATCOM.LEO.NDBPO.\#1>>} by extracting the useful instances from the \textsf{log file} and then enriching them with the information gathered from the \textsf{vector file}. Fig.\ref{A sample of the contents of the <<log>>} file shows how \textsf{Omnet++} generates the contents of the \textsf{<<final log file>>} in Fig.\ref{Figure :Flux de création des fonctionnalités du dataset}.

\begin{figure}[!h]
\begin{center}
\includegraphics[scale=0.55]{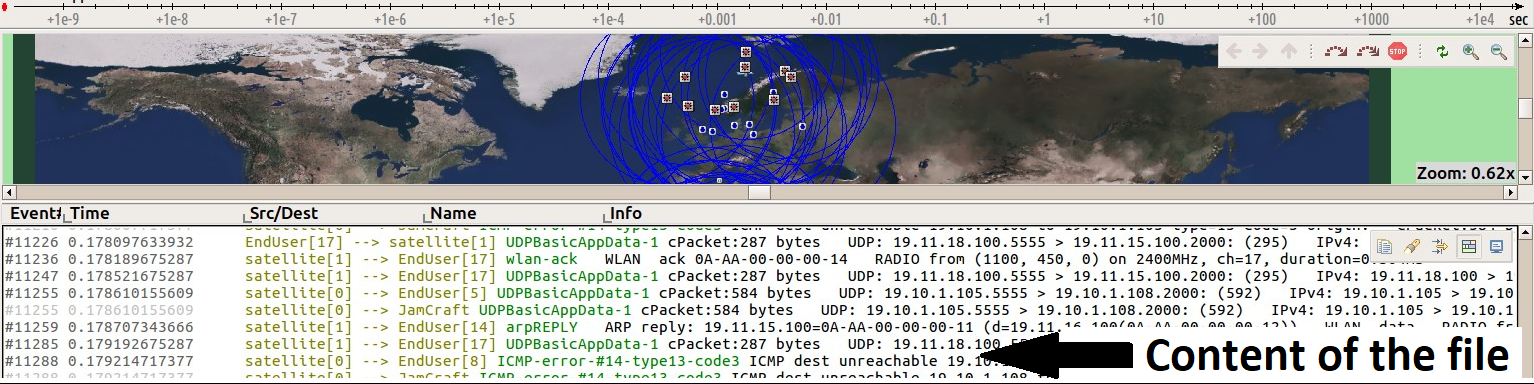}
\caption{ A sample of the contents of the <<log>> file  }
\label{A sample of the contents of the <<log>>}
\end{center}
\end{figure}
For \textsf{<<SATCOM.LEO.NDBPO.\#2>>} we use the pcap file generated by the simulator and captured with \textsf{PcapRecorder} from \textit{satellite[0] node}, then on a new Ubuntu VM equipped with \textsf{<<TcpReplay>>} and a  customized sniffer, that capture packets in real time and simultaneously creates instances of the second dataset.
\subsection{Information on the generated datasets}
The generation time of a dataset is as important as the response time and the detection time, whether offline or in real time. Tab.\ref{time to generate1} illustrates some statistics on the cost in terms of time.

\begin{table}[!h]
 \caption{Statistics on the datasets generation} \vspace{0.3cm}
\footnotesize	
\centering
\begin{tabular}{|c|c|c|c|c|c|c|}
  \cellcolor[HTML]{B2BEB5}Dataset &  \multicolumn{3}{c|}{\cellcolor[HTML]{B2BEB5} SATCOM.LEO.NDBPO.\#1 } &  \multicolumn{3}{c}{\cellcolor[HTML]{B2BEB5} SATCOM.LEO.NDBPO.\#2 }

\\ \hline
 \cellcolor[HTML]{B2BEB5}Scenarios & \cellcolor[HTML]{B2BEB5}  \cellcolor[HTML]{B2BEB5}Scn 1+2  &   \cellcolor[HTML]{B2BEB5}Scn 1 &  \cellcolor[HTML]{B2BEB5}Scn 3 &  \cellcolor[HTML]{B2BEB5}Scn 1+2  &   \cellcolor[HTML]{B2BEB5}Scn 1 &  \cellcolor[HTML]{B2BEB5}Scn 3 \\\hline
Packet extraction time  & 47.70s &  18.92s &  30.88s  & 1500s &   1500s &  1500s \\\hline
Time to remove unnecessary csvs  & 9.28s &  7.83s &  8.43s & / & / & /   \\\hline
Feature creation time  & 1089.49s & 842.53s  &  470.58s & / & / & / \\\hline
Time to add features  & 111.88s & 26.68s  &  10.24s    & 242s & 151s  &  160s \\\hline 

Radio features addition time  & /& /& / & 16h & 9h  &  10h \\\hline 
Normalization time & 
\multicolumn{3}{c|}{ 22.93s} & 
\multicolumn{3}{c|}{ 278.60s}

 \\\hline
\end{tabular}
\label{time to generate1}
\end{table}





Note that two datasets cannot be compared without context, because the first contains more information with a total monitoring on the network in a simulation period of 330s (total~size~=~252MB), and the second dataset is a capture of the satellite[0] during a simulation period of 4500s (total~size~=~929 Mo).

\subsubsection {Dataset with binary classes}
The classes \textsf{<<Normal>>} \& \textsf{<<Rain\_and\_Thunderstorms>>} are~considered as a single class which represents the benign flow, and the two classes \textsf{<<UDP\_Flood\_attack>>} \& \textsf{<<Jamming\_attack>>} are considered  attack flows. Fig.\ref{Figure: the distribution of the flow in binary class1} and Fig.\ref{Figure: the distribution of the flow in binary class2} demonstrate these distributions.

\begin{figure}[!h]
  \centering
  \begin{minipage}[b]{0.48\textwidth}
    \includegraphics[scale=0.66]{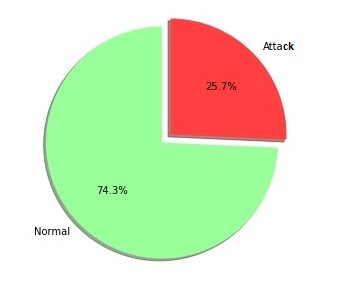}
    \caption{Flow breakdown of \\<<SATCOM.LEO.NDBPO.\#1>>}
    \label{Figure: the distribution of the flow in binary class1}
  \end{minipage}
  \hfill
  \begin{minipage}[b]{0.48\textwidth}
    \includegraphics[scale=0.66]{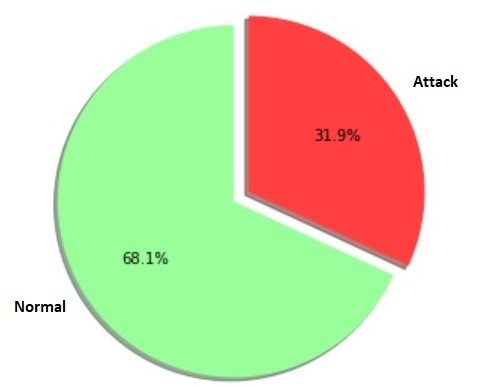}
    \caption{Flow breakdown of \\<<SATCOM.LEO.NDBPO.\#2>> }
  \label{Figure: the distribution of the flow in binary class2}
  \end{minipage}
\end{figure}
\subsubsection {Multi-class dataset}
Each class has its own label. Fig.\ref{Figure: breakdown of flow by type1} and Fig.\ref{Figure: breakdown of flow by type2} demonstrate the flow breakdown by class.

\begin{figure}[!h]
  \centering
  \begin{minipage}[b]{0.48\textwidth}    \includegraphics[scale=0.7]{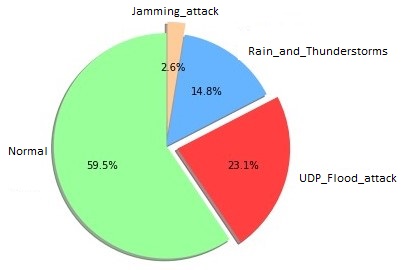}
    \caption{Flow breakdown* of\\ <<SATCOM.LEO.NDBPO.\#1>>}
    \label{Figure: breakdown of flow by type1}
  \end{minipage}
  \hfill
  \begin{minipage}[b]{0.48\textwidth}
    \includegraphics[scale=0.7]{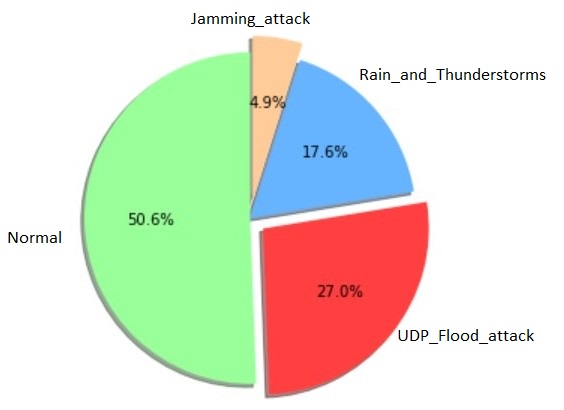}
    \caption{Flow breakdown* of \\<<SATCOM.LEO.NDBPO.\#2>>}
 \label{Figure: breakdown of flow by type2}
  \end{minipage}
\end{figure}

Figs.\ref{Figure: distribution of normal flow and abnormal flow1} and \ref{Figure: distribution of normal flow and abnormal flow2} show more distribution details of normal and abnormal flows of the three interruption types. Both datasets can be found at \url{https://github.com/NacereddineSitouah/Interruption_LEO_SAT_master}.

\begin{figure}[!h]
  \centering
  \begin{minipage}[b]{0.45\textwidth}
    \includegraphics[scale=0.99]{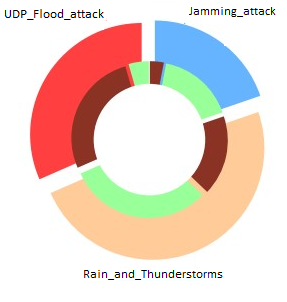}
    \caption{Flow breakdown** of \\ <<SATCOM.LEO.NDBPO.\#1>>}
    \label{Figure: distribution of normal flow and abnormal flow1} 
  \end{minipage}
  \hfill
  \begin{minipage}[b]{0.45\textwidth}
    \includegraphics[scale=0.66]{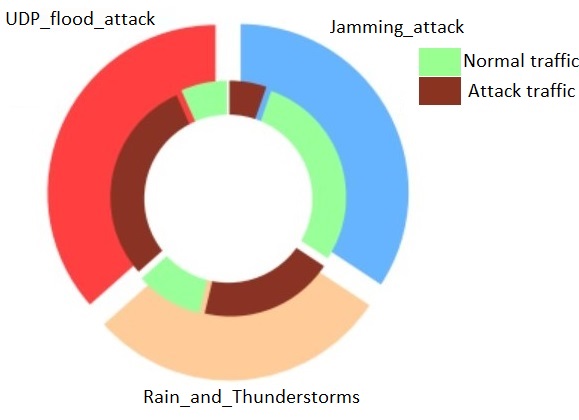}
    \caption{Flow breakdown** of \\ <<SATCOM.LEO.NDBPO.\#2>>}
    \label{Figure: distribution of normal flow and abnormal flow2} 
  \end{minipage}
\end{figure}
\newpage


\section {Performance evaluation of the proposed model
}

In this section, we apply the following algorithms for binary and multi-class classifications: MLP, CNN, RNN, GRU, and LSTM. \\
We use \textsf{Anaconda v4.10.1}\cite{[52]} 
 which is a \textsf{python v3.8.5 distribution} that includes several data-science libraries. We also use the \textsf{Pytorch} \cite{[53]} 
 library, which is one of the best for machine learning. For programming and building models we use \textsf{Jupyter Notebook} \cite{[54]}
and \textsf{VS Code v1.41.1} \cite{[55]}.
\par
Classifiers  parameters : 
Tab.\ref{parametre_des_modeles_pour_datasets} contains the various parameters used in the classification models that had the best performances. The selection procedure was performed by setting multiple parameters and then altering them one by one until the optimal parameters were found. For all models we find that \textit{ReLU}, \textit{Softmax} and the value \textit{0.003} were the best as activation function, output activation function and learning rate respectively, we also used the value 0 as dropout rate through all models.

\scriptsize
\setcounter {LTchunksize} {100}

\begin{longtable}[!h]{|p{6cm}|p{3.8cm}|p{3.8cm}|}
\caption{Deep learning classifier's parameters}

\label{parametre_des_modeles_pour_datasets}
\endfirsthead
\endhead
\hline
\cellcolor[HTML]{B2BEB5}Dataset & \cellcolor[HTML]{B2BEB5}SAT-COM.LEO.NDBPO.\#1 & \cellcolor[HTML]{B2BEB5}SAT-COM.LEO.NDBPO.\#2  \\\hline
\multicolumn{3}{|c|}{\cellcolor[HTML]{F2F3F4} MLP}\\\hline
Input size & 19 & 14  \\
Number of hidden layers & 2 & 4 \\
Number of neurons - hidden layer 1 & 350  & 64 \\
Number of neurons - hidden layer 2 & 400 & 140 \\
Number of neurons - hidden layer 3 & / & 200 \\
Number of neurons - hidden layer 4 & / & 32 \\
Number of epochs & 50 & 200 \\
Batch size & 1  & 300\\
Loss Function & Negative Log Likelihood Loss & Cross Entropy Loss  \\
Optimizer algorithm & Stochastic Gradient Descent Optimizer & adaptive moment estimation (Adam) \\\hline
\multicolumn{3}{|c|}{\cellcolor[HTML]{F2F3F4} <<C>> = RNN / LSTM / GRU }\\\hline
Input size & 19  & 14\\
Number of layers <<C>>   & 3 & 2  \\
Hidden projection layer size (dimension)  &  132 & 132\\
Activation function & Tanh & Tanh \\
Number of epochs (RNN) & 50 & 800 \\
Number of epochs (GRU) & 50 & 200\\
Number of epochs (LSTM) & 50 & 100 \\
Batch size & 1 & 600 \\
Loss Function & Negative Log Likelihood Loss & Negative Log Likelihood Loss  \\
Optimizer algorithm & Stochastic Gradient Descent Optimizer & Stochastic Gradient Descent Optimizer  \\\hline

\multicolumn{3}{|c|}{\cellcolor[HTML]{F2F3F4} CNN}\\\hline
Input size & 18 (3*6) & / \\
Number of convolutional layer & 3  & /\\ 
Number of convolutional neuron - conv layer  1 & 8 & / \\ 
Kernel size / stride / padding  - conv layer 1 & (3x3) / 1 / 1   & /\\ 
 Kernel size / stride - Pooling layer 1 & (1x1) [no effect] / 1  & / \\ 
Number of convolutional neuron- conv layer 2 & 12 & / \\ 
Kernel size / stride / padding  - conv layer 2 & (3x3) / 1 / 1   & / \\ 
Kernel size / stride - Pooling layer 2 & (1x1) [no effect] / 1  & /  \\  
Number of convolutional neuron - conv layer 3 & 18  & / \\ 
Kernel size / stride / padding  - conv layer 3 & (3x3) / 1 / 1   & / \\ 
Kernel size / stride - Pooling layer 3 & (1x1) [no effect] / 1  & / \\ 
Number of fully connected layers & 2  & /\\ 
Number of neurons - hidden layer 1 & 350   & /\\ 
Number of neurons - hidden layer 2 & 400  & / \\ 
Number of epocs & 50  & /\\ 
Batch size & 1 & / \\ 
Loss function & Negative Log Likelihood Loss & / \\ 
Optimizer algorithm & Stochastic Gradient Descent Optimizer  & /\\\hline
\end{longtable}
\normalsize
\subsection {Binary classification}  
Fig.\ref{Figure :  Graphe de perte pour la classification binaire} shows the training loss and validation loss over time for binary classification of~\textsf{<<SATCOM.LEO.NDBPO.\#1>>}, overall, there is always a sharp drop in~the~beginning epochs of learning for both losses in all models, this indicates that all the~models perform very well during at the start, then the learning stops after about ten or~twenty epochs, and this  is either a sign of a local optimum trap, or that the best parameters were found. In addition, we observe that the \textsf{RNN} model starts overfitting on the training data since the training loss starts decreasing rapidly but validation loss maintain almost the same level, that is why this model performs the worst. The~\textsf{CNN} model seems to have reached its peak after the first decade of epochs.

\vspace{-0.5cm}
\begin{figure}[htb]
\begin{center}
\includegraphics[scale=0.57]{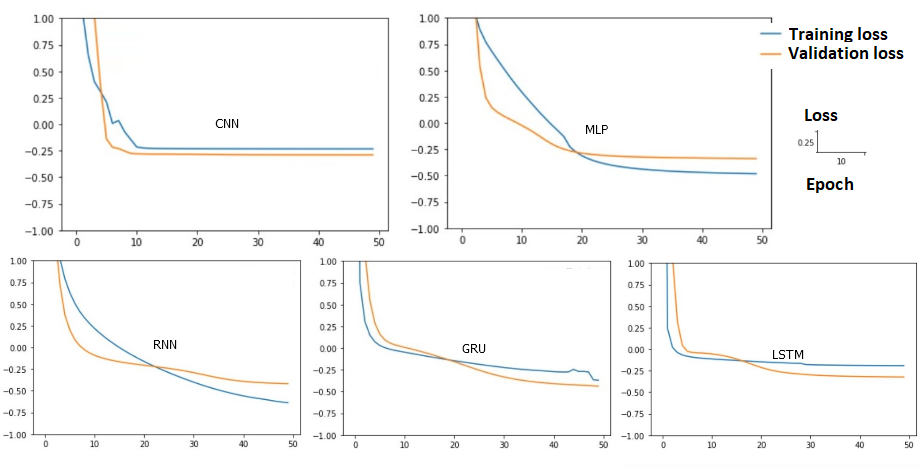}
   \captionsetup{justification=centering}
\caption{Training and validation loss for the binary classification on <<SAT-COM.LEO.NDBPO.\#1>>}
 \label{Figure :  Graphe de perte pour la classification binaire}
\end{center}
\end{figure}

Fig.\ref{Figure: Loss graph for binary classification3} illustrates the graphs for the \textsf{<<SATCOM.LEO.NDBPO.\#2>>} dataset, it shows that almost all the models had similar performances after training, where \textsf{GRU} and~\textsf{LSTM} had a better success in learning, while the other two algorithms \textsf{RNN} and \textsf{MLP} clearly show sign of overfitting  after 500 and 100 epochs, respectively.

\vspace{-0.5cm}
\begin{figure}[htb]
\begin{center} \includegraphics[scale=0.68]{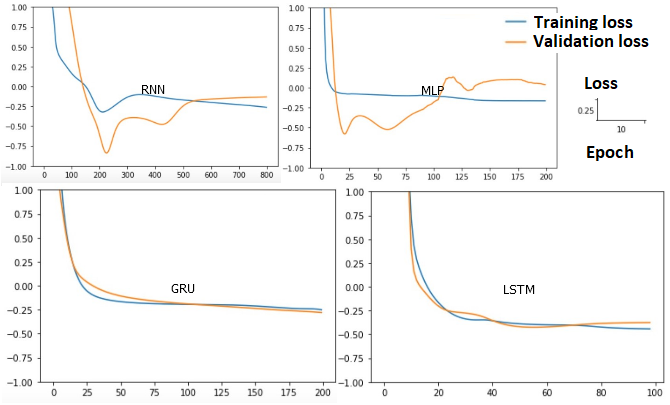}
    \captionsetup{justification=centering}
    \caption{Training and validation loss for the binary classification on <<SAT-COM.LEO.NDBPO.\#2>> }
    \label{Figure: Loss graph for binary classification3}
\end{center}
\end{figure}


Tab.\ref{Performance_results_for_binary_classification} summarizes the final performance results of each model, the \textsf{LSTM} model appears to perform best with a false positive rate of \textit{0.014\%} and a false negative rate of \textit{0.008\%}. However, the CNN model also performs well, but with a false negative rate of \textit{0\%} is not very optimal when the false positive rate is relatively large \textit{0.2\%}, because the false positives themselves will be a new source of disturbance in the network. In addition, the training and execution times are very long which makes the model difficult to update, especially considering that the overall performance of this model is the least-performing one. For this reason, we eliminate  CNN model from the detection experiments on the second dataset.
Tab.\ref{table_de_confusion_binaire} represents the confusion table for binary detection.

\scriptsize
\setcounter {LTchunksize} {100}

\begin{longtable}[!h]{|p{3.2cm}||p{1cm}|p{1cm}|p{1cm}|p{1cm}|p{1cm}||p{1cm}|p{1cm}|p{1cm}|p{1cm}|}
\caption{Performance results for binary classification}

\label{Performance_results_for_binary_classification}
\endfirsthead
\endhead
\hline
\cellcolor[HTML]{B2BEB5}Dataset & \multicolumn{5}{c||}{\cellcolor[HTML]{B2BEB5} SAT-COM.LEO.NDBPO.\#1}  &  \multicolumn{4}{c|}{\cellcolor[HTML]{B2BEB5} SAT-COM.LEO.NDBPO.\#2}  \\\hline
Algorithm & MLP & CNN & RNN  & GRU & LSTM & MLP  & RNN  & GRU & LSTM\\\hline
Accuracy & 99.91\%& 99.85\%& 99.78\%& 99.83\%& 99.98\%& 95.35\%& 95.26\%& 96.12\%& 95.78\%\\\hline
Precision & 99.67\%& 99.46\%& 99.21\%& 99.4\%& 99.96\%& 93.52\%& 93.54\%& 93.39\%& 93.56\%\\\hline
Recall & 99.99\%& 100\%&  99.98\%& 99.99\%& 99.99\%& 93.86\%& 93.61\%& 96\% & 94.98\%\\\hline
Sum of probabilities & 99.91\%& 99.95\%& 99.84\%& 99.86\%& 99.92\%& 98.14\%& 98.61\%& 98.3\%& 98.26\%\\\hline
Training time per epoch & 381.5s & 1474s  & 401.9s & 382.3s & 396.5s & 40.5s & 32.15s & 82.33s & 115.8s\\\hline
Execution time & 67.85s & 166.42s & 136.45s & 124s & 134.39s & 394.5s & 136.45s & 287.45s &  272.65s\\\hline

\end{longtable}

\begin{table}[htb]
\caption{Table de confusion - Classification binaire}
\label{table_de_confusion_binaire}
\scriptsize
\centering
\begin{tabular}{|p{1.15cm}|p{1cm}||p{1cm}|p{1cm}|p{1cm}|p{1cm}|p{1cm}|p{1cm}|p{1cm}|p{1cm}|p{1.08cm}|p{1cm}|}
\hline\hline

\multicolumn{2}{|c||}{\cellcolor[HTML]{B2BEB5}Algorithm} & \multicolumn{2}{c|}{\cellcolor[HTML]{B2BEB5}MLP} &  \multicolumn{2}{c|}{\cellcolor[HTML]{B2BEB5}CNN} &  \multicolumn{2}{c|}{\cellcolor[HTML]{B2BEB5}RNN}  &  \multicolumn{2}{c|}{\cellcolor[HTML]{B2BEB5}GRU} &  \multicolumn{2}{c|}{\cellcolor[HTML]{B2BEB5}LSTM} \\\hline
Dataset &Class &Normal & Attack  &Normal & Attack  &Normal & Attack  &Normal & Attack  &Normal & Attack  \\\hline
 \multirow{2}{*}{D-Set\#1} & Normal & 99.87\% & 0.01\%& 99.79\%& 0\%& 99.69\%& 0.02\%& 99.76\%& 0.01\%&99.985\%& 0.009\%\\
 & Attack & 0.13\%& 99.99\%& 0.21\%& 100\%& 0.3\%& 99.98\%& 0.23\%& 99.99\%& 0.01\%& 99.99\%\\
    \hline
 \multirow{2}{*}{D-Set\#2}& Normal & 96.35\% & 6.54\%& / & / & 96.36\%& 6.82\%&  96.27\%& 4.17\%& 96.35\%& 5.26\%\\
  & Attack  & 3.65\%& 93.46\%& / & / & 3.64\%& 93.18\%& 3.73\%& 95.83\%& 3.65\%& 94.71\%\\
    \hline
\end{tabular}
\end{table}


\normalsize
\subsection {Multi-class classification}
Fig.\ref{Figure :  Graphe de perte pour la classification multiclasse} represents the loss graph for the dataset <<SAT-COM.LEO.NDBPO.\#1>> in~multiclass classification. The CNN graph looks very unstable, inconsistent and doesn't converge smoothly. In spite of that, RNN and LSTM reach their peak since both graphs remain steady after few epochs. On the other hand MLP and GRU seem to perform well and can even continue learning if granted more training epochs.

\begin{figure}[htb]
\begin{center}
 \includegraphics[scale=0.56]{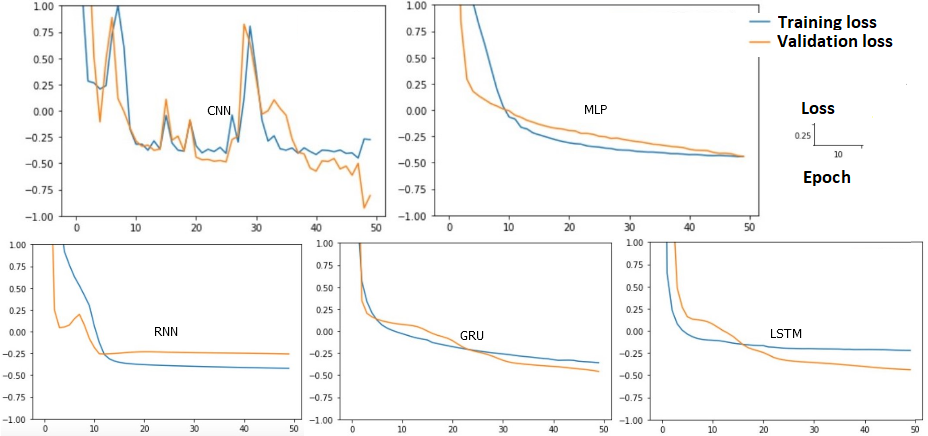}
    \captionsetup{justification=centering}
\caption{Training and validation loss for the multiclass classification on <<SAT-COM.LEO.NDBPO.\#1>>}
 \label{Figure :  Graphe de perte pour la classification multiclasse}
\end{center}
\end{figure}

Fig.\ref{Figure: Loss graph for multiclasse classification3} shows results for \textsf{<<SAT-COM.LEO.NDBPO.\#2>>}, we can see that there is no over-fitting for all models, we also observe that the decrease in the losses is smoother until they stabilize when the models reach their peaks.
\begin{figure}[htb]
\begin{center}    \includegraphics[scale=0.67]{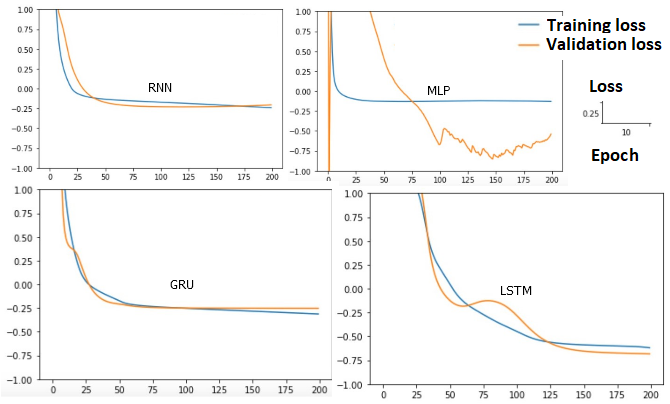}
\captionsetup{justification=centering}
    \caption{Training and validation loss for the multiclass classification on <<SAT-COM.LEO.NDBPO.\#2>> }
    \label{Figure: Loss graph for multiclasse classification3}
\end{center}
\end{figure}


Tab.\ref{Temps et performance4} summarizes the final performance of each model for the multiclass classification. \textsf{MLP} outperforms other algorithms with a false positives rate of \textit{0.02\%} and a false negatives rate of \textit{0.02\%} as well. \textsf{RNN} had a false positives rate and~false negatives of \textit{0\%} since all false classifications were classified as udp flood attack. The~\textsf{CNN} confirms that it does not adapt well to the detection of~interruptions in~the networks. The table shows for the second dataset, that MLP performs better, mainly due to the relatively low false positives rate.
The~performance shown in Table 4.11 indicates that MLP performs better, mainly due to the relatively low false positive rate (12,25~\%)

\scriptsize
\setcounter {LTchunksize} {100}

\begin{longtable}[!h]{|p{3.2cm}||p{1cm}|p{1cm}|p{1cm}|p{1cm}|p{1cm}||p{1cm}|p{1cm}|p{1cm}|p{1cm}|}
\caption{Performance results for multiclass classification}

\label{Temps et performance4}
\endfirsthead
\endhead
\hline
\cellcolor[HTML]{B2BEB5} Dataset & \multicolumn{5}{c||}{\cellcolor[HTML]{B2BEB5} SAT-COM.LEO.NDBPO.\#1}  &  \multicolumn{4}{c|}{\cellcolor[HTML]{B2BEB5} SAT-COM.LEO.NDBPO.\#2}  \\\hline
Algorithm & MLP & CNN & RNN  & GRU & LSTM & MLP  & RNN  & GRU & LSTM\\\hline
Accuracy & 99.33\%& 97.32\%& 96.02\%& 98.69\%& 98.72\%& 90.94\%& 94.35\%& 94.35\%& 94.35\%\\\hline
Precision & 99.98\%& 99.9\%& 100\%& 99.84\%& 99.59\%& 87.75\%& 85.94\%& 85.94\%& 85.94\%\\\hline
Recall & 99.98\%& 99.9\%& 100\%& 100\%& 99.99\%& 71.39\%& 100\%& 100\%& 100\%\\\hline
Sum of probabilities & 99.7\%& 99.07\%& 99.94\%& 99.82\%& 99.65\%& 99.66\%& 99.48\%& 99.55\%& 99.12\%\\\hline
Training time per epoch & 404.54s & 1770.6s & 367.32s & 382.69s & 414.62s & 31.77s & 26.89s & 77.18s & 119.93s\\\hline
Execution time & 104.88s & 157.95s & 130.29s & 129.72s & 138.55s & 414.95s & 238.18s & 276.46s & 372.82s\\\hline

\end{longtable}
\normalsize
From Tab.\ref{table de confusion multiclass}, we can deduce that the most efficient models to minimize the false positives rate are \textsf{MLP} and \textsf{GRU} with the optimizer algorithm <<Adam>>. More precisely \textsf{MLP} is ideal for the detection of \textsf{<<DDoS\_UDP>>} attacks despite the low detection rate which is \textit{83~\%}, because detecting certain attack packets is sufficient to detect malicious communications.
On the other hand, the \textsf{GRU} is more efficient at detecting jamming attacks because less false positives are classified with Jamming\_attack category. 
\begin{sidewaystable}
\caption{Confusion table - Classification multiclass}
\label{table de confusion multiclass}
\scriptsize	
\centering

\begin{tabular}{|p{1.45cm}|p{2cm}||p{1.2cm}|p{1.7cm}|p{2cm}|p{1.2cm}||p{1.2cm}|p{1.7cm}|p{2cm}|p{1.3cm}|}

\hline\hline
\multicolumn{2}{|c||}{\cellcolor[HTML]{B2BEB5}Algorithm} & \multicolumn{4}{c||}{\cellcolor[HTML]{B2BEB5}MLP} &  \multicolumn{4}{c}{\cellcolor[HTML]{B2BEB5}RNN}  \\\hline
Dataset &Class &Normal & UDP\_Flood  & Rain\_Thunder & Jamming &Normal & UDP\_Flood  & Rain\_Thunder & Jamming     \\\hline
 \multirow{4}{*}{Dataset\#1} & Normal & 98.82\%& 0.02\%& 0\%& 0\%  & 92.95\%& 0\%& 0\%& 0\%    \\
 & UDP\_Flood & 0\%& 99.98\%& 0\%& 0\%  & 0\%& 100\%& 0\%& 0\%      \\
 & Rain\_Thunder & 1.16\%& 0\%& 100\%& 0\%  & 7.05\%& 0\%& 100\%& 0\%    \\
 & Jamming & 0.0099\%& 0\%& 0\%& 100\%  & 0\%& 0\%& 0\%& 100\%    \\
    \hline
 \multirow{4}{*}{Dataset\#2} & Normal & 93.7\%& 93.7\%& 0\%& 0\%  & 89.659\%& 0\%& 0\%& 0\%   \\
 & UDP\_Flood & 0.0006\%& 83.54\%& 0\%& 0.0065\%  & 4.0913\%& 100\%& 0\%& 0\%   \\
 & Rain\_Thunder & 0\%& 0\%& 100\%& 0\%  & 0\%& 0\%& 100\%& 0\%  \\
 & Jamming & 6.294\%& 0\%& 0\%& 99.993\%  & 6.2494\%& 0\%& 0\%& 100\% \\
    \hline\hline
 \multicolumn{2}{|c||}{\cellcolor[HTML]{B2BEB5}Algorithm} &  \multicolumn{4}{c||}{\cellcolor[HTML]{B2BEB5}GRU} &  \multicolumn{4}{c}{\cellcolor[HTML]{B2BEB5}LSTM}  \\\hline
Dataset &Class &Normal & UDP\_Flood  & Rain\_Thunder & Jamming & Normal & UDP\_Flood  & Rain\_Thunder & Jamming     \\\hline
 \multirow{4}{*}{Dataset\#1} & Normal & 97.68\%& 0\%& 0\%& 0\%  & 97.73\%& 0.0084\%& 0.022\%& 0\%    \\
 & UDP\_Flood & 0.033\%& 100\%& 0\%& 0\%  & 0\%& 99.99\%& 0\%& 0\%      \\
 & Rain\_Thunder & 2.24\%& 0\%& 100\%& 0\%  & 2.065\%& 99.71\%& 99.98\%& 0\%    \\
 & Jamming & 0.043\%& 0\%& 0\%& 100\%  & 0.2\%& 0\%& 0\%& 100\%    \\
    \hline
 \multirow{4}{*}{Dataset\#2} & Normal & 92.58\%& 0\%& 0\%& 1.142\%  & 89.66\%& 0\%& 0\%& 0\%  \\
 & UDP\_Flood & 4.14\%& 100\%& 0\%& 0\%  & 4.09\%& 100\%& 0\%& 0\%   \\
 & Rain\_Thunder & 0\%& 0\%& 100\%& 0\%  & 0\%& 0\%& 100\%& 0\%  \\
 & Jamming & 3.441\%& 0\%& 0\%& 98.85\%  & 6.2494\%& 0\%& 0\%& 100\% \\
    \hline
\end{tabular}
\end{sidewaystable}
\end{itemize}


\newpage
\normalsize
\subsection {Realtime detection}
To perform the real-time detection, we programmed  another sniffer  that captures the traffic for a given period of time, and at the end of each period the sniffer executes a code that processes the captured packets, loads the models and launches the process of attack detection with the captured traffic. The discovery procedure is performed without interrupting the continuous capture of the network.
\par We program a real-time IDS which operates in two modes. \\1) Normal Mode: In this mode, we try to eliminate false positives completely at the price of losing some precision. \textit{i.e} our {IDS} does not try to decide for each suspect packet if it is malicious or victim, but it only  detects anomalies in a communication flow.
2) {Safe mode:}
In this mode, the {IDS} detects as many suspicious packets as possible, and does not give as much importance to false positives as to false negatives. This model should be implemented passively, and it should never be a proactive system as it will alert on many false positives.
\\For real-time simulation, we take the pcap file generated with \textsf{Omnet++} and~we~reproduce the traffic using Tcpreplay. We also use the  extracted information  from the simulation on the \textit{SNIR} and on the \textit{THROUGHPUT} of~the~satellite during communication, since these two features are available and~can be calculated in all wireless networks. 
The detection process does not differ too much from offline detection, a continuous discovery  captures packets and~processes them in very short periods of time. To minimize the false positive rate, the~\textit{<<Normal>>} mode detection is as follows:
\begin {itemize} [leftmargin = -0.35em]
    \item [\textbullet] \textsf{UDP Flood}: Throws alert if both of the following conditions are true: \\ - \textit{20\%} minimum of captured traffic is classified as a \textit{<<flood>>} attack. \\
    - At least \textit{70\%} of a captured flow is a \textit{<<flood>>} class attack.
    \item [\textbullet] \textsf{Natural Phenomenons}: Throws alert if both of the following conditions are true: \\
    - A maximum of \textit{10\%} of the captured traffic is classified as a \textit{<<Rain and Thunderstorms>>} attack. \\
    - \textit{20\%} of a captured stream is affected by \textit{<<Natural interference>>}.
    \item [\textbullet] \textsf{Jammed traffic}: Throws alert if both of the following conditions are true: \\
    - A maximum of \textit{20\%} of captured traffic is classified as a \textit{<<Network Jamming>>} attack. \\
    - \textit{90\%} minimum of a captured traffic is affected by \textit{<<Network Jamming>>}.
\end {itemize}

\begin{figure}[!h]
\begin{center}
\includegraphics[scale=0.8]{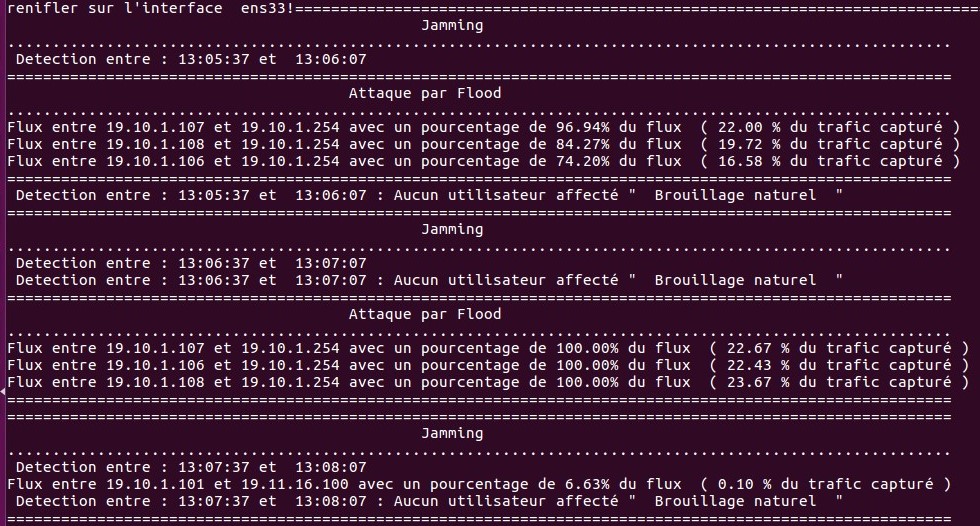}
\captionsetup{justification=centering}
\caption{Behavior of IDS in <<Normal>> mode against malicious "UDP Flood". }
\label{Behavior of IDS in <<Normal>> mode against malicious "UDP Flood".}
\end{center}
\end{figure}

\begin{figure}[!h]
\begin{center}
\includegraphics[scale=0.65]{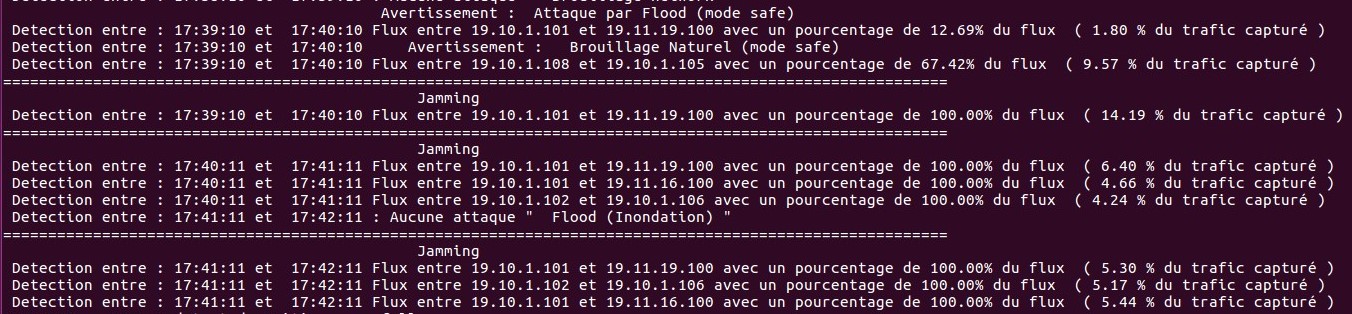}
\captionsetup{justification=centering}
\caption{Behavior of IDS in <<Safe>> against  << Jamming >> Attacks.}
\label{Behavior of the IDS in <<Safe>> mode in the face of malicious attacks "Network jamming".}
\end{center}
\end{figure}

\textit{The IDS in action:} 
In the  following we  show a comparison of detection in \textit{<<Normal>>} mode and in \textit{<<safe>>} mode for the three existing classes. The {IDS} is~executed with a~period of \textit{30 seconds}. The results for the first \textit{2 minutes} are as follows:\\
For benign traffic, the \textit{Normal} mode do not trigger any alert for all the three attacks. Furthermore, 
 in \textit{Safe} mode no alert has been triggered for the \textsf{Rain\_and\_Thunderstorms} class, but on the other hand, for \textsf{Jamming} and~\textsf{UDP\_Flood} class, several alerts are launched, the normal traffic classified as flood attack is on average between \textit{2\%} and~\textit{6\%}, while for normal packets classified as victims of jamming attacks is between \textit{21\%} and \textit{28\%}.
The results of detection in~both modes for the malicious flood traffic shows that no packet is misclassified as~a~victim of natural interference, while a~small flow composing \textit{0.1\%} is miss-classified as a jamming victim, all malicious flows of flood attacks seem to be well classified. Fig.\ref{Behavior of IDS in <<Normal>> mode against malicious "UDP Flood".} illustrates an example of the flood attacks detection in \textit{Normal} mode.
\\For traffic affected by natural phenomena events, results in both modes are  \textit{100\%} accurate, although in \text{Safe} mode, \textit{0.46\%} of a normal flow, that translates to about \textit{0.04\%} of the total traffic captured,  is falsely classified as affected. Same goes for~the detection of jamming traffic in \textit{Normal} mode, the malicious traffic is~successfully detected, but on the other hand in \textit{Safe} mode, two flows are detected and classified as \textsf{UDP\_Flood} and  \textsf{Rain\_and\_Thunderstorms} with \textit{12.69\%} and \textit{67.42\%} of~total flow of the two classes respectively, as presented is Fig.\ref{Behavior of the IDS in <<Safe>> mode in the face of malicious attacks "Network jamming".}.


\newpage
\section{Conclusion and future work}
An interruption detection method is proposed for LEO satellite networks based on Deep Learning in this paper. We have studied the various vulnerabilities and~risks that threaten the operation and fluidity of Satellite communications. This~work focuses on disruption threats of different types, such as flood attacks, network jamming attacks and natural phenomena events (such as rains, thunderstorms and hurricanes). 
 Several deep learning based interruption detection models are proposed on two of our own generated datasets, including models based on MLP, CNN, RNN, GRU and LSTM. We have also provided performance assessment for binary and multi-class classifications. The best recorded accuracy on the dataset SATCOM.LEO.NDBPO.\#1 is 99.987\% for binary traffic detection and 99.33\% for multiclass traffic detection with a minimum false positive rate of~0.0144\%. For the dataset  SATCOM.LEO.NDBPO.\#2, the best recorded accuracy  is 96.12\% for the detection in binary classification and 94.35\% for~multiclass classification with a~minimum false positives rate of 3.72\% using a~hybrid model composed of MLP and GRU. Enhancing this hybrid model with some statistical constraints also improves detection results and decreases false positives rate to almost 0\%, at the cost of losing some  accuracy precision.
In~conclusion, the results shows that this approach is effective for both offline and online interruption attacks detection, based on accuracy rate, detection rate, and prediction speed. This helps to demonstrate that Deep Learning algorithms could  very well improve the  security of satellite networks, especially against DDoS attacks. 
As future work, and under better circumstances such as, more computing power and memory resources, this work can be continued by implementing the DTN (Delay/Disruption Tolerant Networking) protocol from NaSa \cite{[56]}
which, according to NaSa, is perhaps the future of satellite communication. This protocol is currently being tested and improved. Different deep learning methods can can have better accuracy rate by involving  terrestrial satellite terminals in the detection process, such as FL (Federated learning) \cite{[57]} 
to protect users privacy and enable their contributions to learning, combined with DAEs and other promising models such as Bi-LSTM to speed up turnaround time.


%




\ifCLASSOPTIONcaptionsoff
  \newpage
\fi



\bibliographystyle{IEEEtranTCOM}
\end{document}